# Extracting relevant structures from self-determination theory questionnaires via Information Bottleneck method

Daniela M. L. Barbato[1] and Jean-Jacques De Groote[2]

[1]Instituto SEB de Educação, Rua Deolinda 70, Jardim Macedo, Ribeirão Preto, 14091-018, SP, Brazil

[2]Centro Universitário Moura Lacerda, Av. Dr. Oscár de Moura Lacerda, 1520 - Independência, Ribeirão Preto - SP, 14076-510 SP, Brazil

**Abstract**

In this paper, we introduce the application of Information Bottleneck as a method to investigate properties of questionnaires developed for the study of motivational profiles based on self-determination theory. Founded on information theory, the Information Bottleneck method compresses multidimensional categorical data into clusters by minimizing the loss of relevant information. It does not require linearity such as Pearson correlation. For exploratory data analysis, it allows investigating the structure of questionnaire items grouping with a hierarchical partition as perceived by the sample, based only on conditional probability distributions calculated from answers to the questionnaire items. We applied the approach to an instrument adapted from the academic motivation scale. The Information Bottleneck approach showed a partition of items aligned with the self-determination continuum. The results of this first application of Information Bottleneck to psychometric studies suggested that this procedure can could be useful as a consistent, complementary analysis for traditional exploratory procedures such as factor analysis.

# Introduction

The classical approach to study motivation considers the distinction between two main types. The intrinsic motivation, inherent in the individual, relates to personal satisfaction, and extrinsic motivation, individual do something to obtain separable outcome, as avoiding punishment, or gain recognition and reward (Deci & Ryan, 1985).

In self-determination theory (SDT) proposed by Deci and Ryan in 2000, human motivation is described as a continuum. At one extreme is the amotivation, or lack of motivation, at the other is the intrinsic motivation. Between these is extrinsic motivation, divided into four levels related to the reason why an individual does a task: External regulation, introjected regulation, identified regulation, integrated regulation. An individual which acts to satisfy an external demand, to receive a reward is motivated with external regulation, an individual which acts to avoid guilt or to maintain self-steam is motivated with introjected regulation.

Self-determination increases as the actions of the individuals become less externally regulated. When an individual does something because identifies the value of an activity to achieve the benefits, he is motivated with identified regulation. When as individual does something because identifies the value of an activity to achieve the benefits, he is motivated with identified regulation. When individual integrates the values of actions to his own values, but acts interested in some outcome, he is motivated with integrated regulation, the most autonomous regulation.

Motivation of individuals in different activities is widely investigated within the theory of self-determination. Quantitative studies in different fields are performed through questionnaires whose items are associated with motivational states. Related to healthy behaving there are motivations to stop smoking (Williams et al., 2006), weight loss (Teixeira, Silva, Mata, Palmeira & Markland, 2012), weight management (Ng, Ntoumanis & Thøgersen-Ntoumani, 2014), for sports practice (Vlachopoulos, Karageorghis & Terry, 2000; Pelletier et al., 1995; Vallerand & Losier, 1999) and physical activities (Cox, Smith & Williams, 2008; Kilpatrick, Hebert & Bartholomew, 2005). Vallerand et al. (1992) developed and validated an instrument to study motivation for learning, the Academic Motivational Scale (AMS). Instruments to investigate motivation in learning science (Glynn, Brickman, Armstrong & Taasoobshirazi, 2011), mathematics (Wæge, 2009, Sullivan & McDonough, 2007) and foreign language (Noels, Pelletier, Clément & Vallerand, 2003; Papi & Teimouri, 2014; Kimura, Nakata & Okumura, 2001) also has been investigated.

One of the main objectives in psychometric research is the construction and validation of instruments, based on theories in the psychology field. Exploratory factor analysis (EFA) is indicated to analyze data from non-validated questionnaires (Glynn, Taasoobshirazi & Brickman, 2009). The goal of this multivariate method of analysis is to find latent variables, or constructs, underlying observed variables. It is useful to reduce the items of a questionnaire into unobserved variables, referred as factors. Despite being an efficient method for motivation studies, results from EFA analysis may be opened to arbitrary decisions made by the researcher (Cattell, 2012). It is suggested for such cases that different techniques be applied in the investigation of psychometric properties of non-validated instruments.

Our goal in this study is to introduce an approach based on information theory, the Information Bottleneck (IB) method (Tishby, Pereira & Bialek, 2000) to investigate psychometric properties of instruments developed for the study of motivational profiles. It allows questionnaires items clustering without the need to establish similarity among questions answers, such as Euclidean distance or correlation. Calculations are made using only the joint probability distribution calculated from the questionnaire answers data. Applications of IB method have been investigated with interesting results in different areas such as speech recognition (Hecht et al., 2009), detection of acoustic events (Wang et al., 2017), documents unsupervised clustering, (Slonim & Tishby, 2000), gene expression data (Tishby & Slonim, 2001) and galaxy spectra classification (Slonim, Somerville, Tishby & Lahav, 2001).

In this study, we apply the IB method for an adaptation of the AMS that has been under exploratory analysis (Guimarães & Bzuneck, 2008; Leal, Miranda & Carmo, 2013). The investigation of how the items of the questionnaire are grouped in motivational states was carried out by means of a gradual process of partition.

In the next section will be described the instrument and the IB approach used in this study.

## Method

**Data and instrument**

Academic motivation scale is an instrument based on the self-determination theory (SDT) proposed by Deci and Ryan (1985). First developed in French, named Echelle de Motivation en Education (EME) (Vallerand, Blais, Brière & Pelletier, 1989), was tested with Canadian students. The results obtained from confirmatory factor analysis produced a seven-factor structure with satisfactory indices of internal consistency and high temporal stability indices. Vallerand et al. (1992) translated it to English, themed Academic Motivational Scale (AMS). Results from this study showed a great similarity between AMS and EME. Recently Guay, Morin, Litalien, Valois and Vallerand (2015) confirmed the construct validity of AMS by using exploratory structural equation modeling. This instrument is composed of 28 items, distributed in seven subscales: amotivation, external regulation, introjected regulation, identified regulation and intrinsic motivation, divided in intrinsic motivation for knowledge, intrinsic motivation toward accomplishments, and intrinsic motivation for stimulation. It does not include integrated regulation items.

Due to the good results obtained, this scale has been widely applied. Fairchild, Horst, Finney and Barron (2005) obtained adequate internal consistence and a well-fitted seven-factor structure with a sample of 1406 American college students. Adequate internal consistence for seven-factor structure was obtained by Cokley, Bernard, Cunningham and Motoike (2001) in a study of a sample composed of different ethnic and gender groups of American students and by Smith, Davy and Rosenberg (2010) with a sample of business students. A translated version of this scale was also studied such as in Spanish (Alonso, Lucas & Izquierdo, 2005), Greek (Barkoukis, Tsorbatzoudis, Grouios & Sideridis, 2008) and Norwegian (Utvær and Haugan, 2016).

The instrument studied in this work is an adaption from AMS as proposed by Guimarães and Bzuneck (2008). They introduced four items related to Integrated regulation, new items for external and identified regulation, and reducing the assertions of intrinsic motivation considering only items that said positive emotions on the student. This instrument with 31 items was applied to 388 students of university in the South of Brazil. Exploratory factor analysis (EFA) furnished a seven-factor structure, two items were discarded and, items related to external regulation divided into two factors, one termed by authors as external regulation by attendance and other as external regulation by social interaction. Some items did not load in the expected factor. The items of questionnaire (Leal et al., 2013) applied in this study is shown in the appendix. The question "why do I come to university?" followed by 29 items in a 7-point Likert scale ranging from 1 = totally disagree to 7 = totally agree, were applied in 2015 during class time to 602 students, who attend a university, coming from a large geographic area within the district.

The students were invited to respond and communicated that their identities would not be revealed. Our sample is composed of engineering (307), accounting (160) and administration (135) students, aged 18 to 30 years, with 24% women and 76% men in engineering, 68% women and 32% men in accounting and 50% women and 50% men in administration courses.

The study proposal was submitted and approved by Brazilian Ethics Committee.

**Information Bottleneck**

Information Bottleneck is a method based on information theory developed by Tishby et al. (2000). By means of mutual information (MI) concept, it allows information compression while preserving its relevance. MI gives a relatedness between random categorical variables $X$ and $Y$ based on the joint probabilities $p(x,y)$ distribution without any assumption of underlying relationship among the variables (Seok & Kang, 2015). It is defined as

$$I(X;Y) = \sum_i \sum_j p(x_i, y_j) \ln \frac{p(x_i, y_j)}{p(x_i)p(y_j)}. \tag{1}$$

If variables are not dependent, MI is null, it can be interpreted as the stored information in one variable $X$ about another $Y$, or the predictability of one variable by knowing other one (Li, 1990).

Under information bottleneck approach the items of the questionnaire represented by variable $X$ are compressed into a lower dimensional variable $T$, in such a way to keep maximum information about variable $Y$, represented by the answers to the questionnaire.

Compression and relevancy are measured by $I(X;T)$ and $I(Y;T)$, respectively. The degree of compression represented by parameter $t$ indicates the number of clusters formed. IB method furnishes the identification of an optimal solution for each value of $t$ by minimizing the functional

$$L[p(t|x)] = I(T;X) - \beta\, I(T;Y), \tag{2}$$

where the Lagrange parameter $\beta$, mediates compression and relevance.

For motivation profiles within Self Determination Theory, IB method is used to partition the items of the questionnaire into clusters. The joint probability distribution of items and answers are obtained from scores for each person on the items.

The algorithm used is this work follows the heuristic process developed by Tishby et al. (2000), with $p(x, y)$, $\beta$ and $t$ as input. An initialization of $p(t\,|\,x)$ is set with random values for the following iterative equations,

$$p(t_i) = \sum_j p(x_j) p(t_i\,|\,x_j) \qquad (3)$$

$$p(y_i\,|\,t_j) = \sum_k p(y_i\,|\,x_k) p(x_k\,|\,t_j) \qquad (4)$$

$$p(t_i|x_j) = \frac{p(t_i)}{Z(x_j,\beta)} exp\left(-\beta \sum_k p(y_k|x_j) \ln \frac{p(y_k|x_j)}{p(y_k|t_i)}\right), \qquad (5)$$

where $Z$ is the partition function defined as,

$$Z(x_j,\beta) = \sum_i p(t_i) exp\left(-\beta \sum_k p(y_k|x_j) \ln \frac{p(y_k|x_j)}{p(y_k|t_i)}\right) \qquad (6)$$

This algorithm finds the probability distributions $p(t\,|\,x)$ and $p(y\,|\,t)$, that shows how variables are grouped, i.e., which variables belong to each cluster. The algorithm should be applied for a set of different initial conditions, as it finds a local minimum solution. However, the best solution can be identified by choosing the one with the lowest $L[p(t\,|\,x)]$.

## Results and Discussion

The instrument chosen in this work for IB application was also analyzed by means of an exploratory factor analysis by Guimarães and Bzuneck (2008) and Leal et al. (2013). They found seven factors, and both analyses indicated items that did not load on the expected factor, shown in appendix.

8The EFA of our data also indicated seven factors as shown in table 1, associated to eigenvalues above 1 that explained 55.9 % of the total data variability. The internal consistency of the questionnaire items was verified through the calculation of the Cronbach's alpha, that yielded α = 0.762. This parameter is limited between 0 and 1, and the closer to 1, the greater the reliability of the instrument. Values greater than 0.7 are considered acceptable (Cortina, 1993).

The Bartlett's sphericity test was applied to reject the null hypothesis of the identity matrix. It provided a significance level of less than 0.03, and the value provided by the Kaiser-Meyer-Olkin (KMO) test to measure the sample adequacy was 0.868, considered admirable (Tabachnick & Fidell, 2007). Table 1 shows how the items are grouped by EFA and the internal consistency of each factor.

Table 1: Factors and their corresponding Cronbach's alpha obtained from EFA.

| Factors | Items | α |
|---|---|---|
| $F_1$ | 1 7 9 13 16 19 | 0.804 |
| $F_2$ | **2** 3 11 | 0.736 |
| $F_3$ | 6 28 29 | 0.464 |
| $F_4$ | 5 8 10 **12 14** 15 20 | 0.645 |
| $F_5$ | 22 **23** 24 25 | 0.558 |
| $F_6$ | 18 26 27 | 0.668 |
| $F_7$ | 4 17 21 | 0.717 |

As can be seen in Table 1, the EFA of the data also indicate items not loading on expected factors (items 2, 12, 14, 23). In fact, the same questionnaire applied to distinct population can yield different factor structure (Massidda, Carta & Altoè, 2016). The EFA is limited to identify a number of latent factors (items grouping) bounded to an explained variance minimum, for our sample seven factors.

We investigated the structure of items grouping through IB method for different number of clusters (*t*). The dendrogram presented in Fig. 1 shows how clusters evolve according to items partition. As proposed by IB method, each cluster structure with respect to *t*, $C_i(t)$, {i = 1...*t*}, corresponds to the configuration with the lowest information loss about student answers.

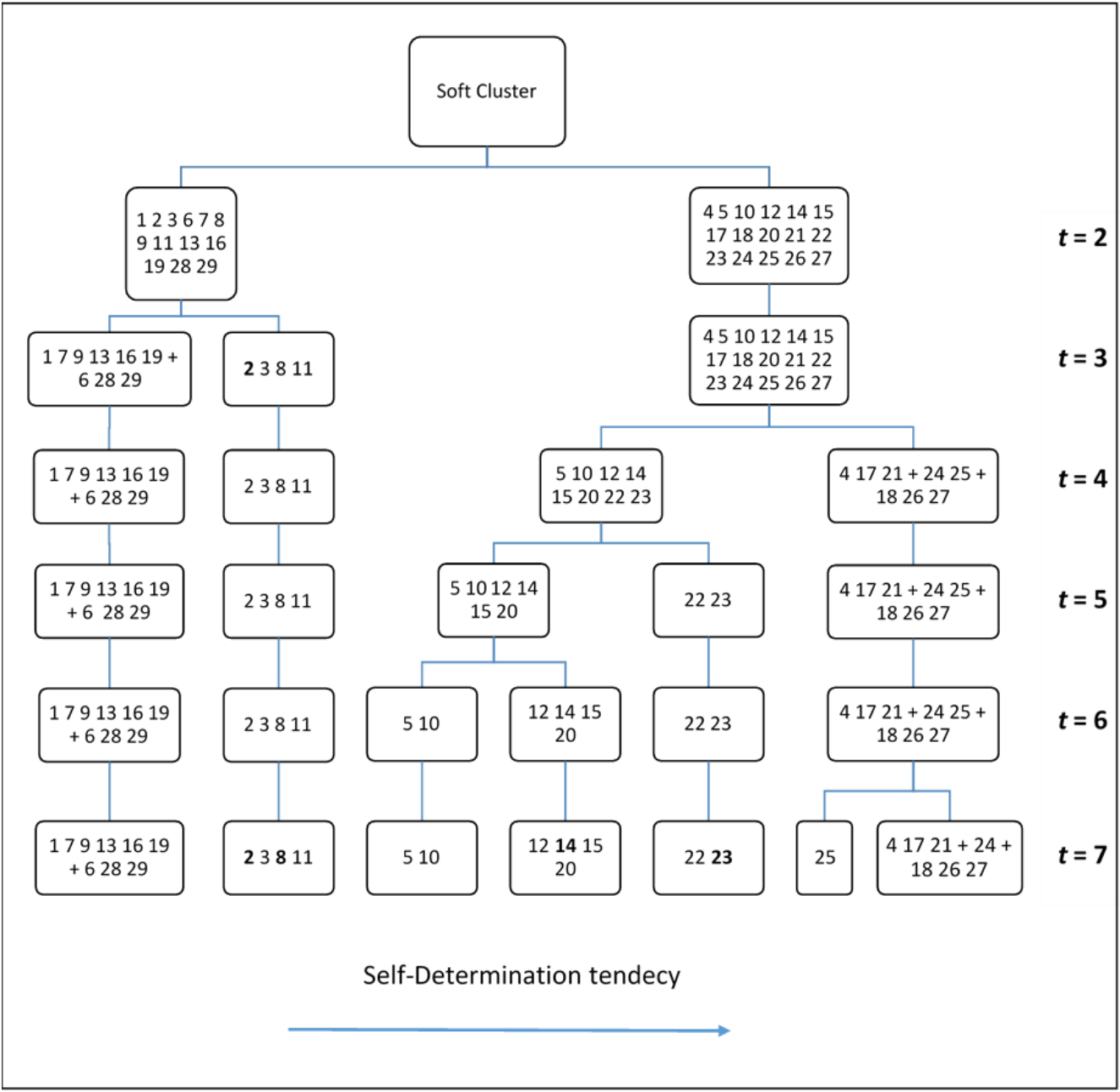

Figure 1: Items clusters partition obtained with IB procedure.

Compressing the information into two clusters we found that cluster $C_1(t=2)$ groups items associated with amotivation and external regulation, except item 2 (identified regulation) and 8 (introjected regulation). Within the second cluster, $C_2(t = 2)$, the introjected, identified and integrated regulations are grouped with the intrinsic motivation, except item 14 (external regulation by attendance). This configuration suggests the classical distinction between intrinsic and extrinsic motivation as pointed out by Ryan and Deci (2000).

When items are compressed into three clusters, the first one, $C_1(t = 2)$, bifurcates into two parts, with the cluster, $C_3(t = 3)$, formed by items {2,3,8,11}, that remain within the same cluster up to $t = 7$. The second cluster stays unchanged: $C_2(t = 2) = C_2(t = 3)$. If items are compressed into four clusters ($t = 4$), the second cluster bifurcates, $C_3(t = 3) \rightarrow C_3(t = 4) + C_4(t = 4)$ so that $C_3(t = 4)$ contains most statements related to introjected regulation, except 12 (integrated regulation), 14 (external regulation by attendance) and 22 (identified regulation), and the other concentrates items related to the more self-determined forms. If the information is compressed into 5 clusters, the only change is a split of the cluster that contains the introjected regulation. When 6 clusters are set, this cluster continues to split as show in Fig.1. Only with seven clusters there is a separation of item {25}, that belongs to identified regulation, from $C_7(t = 7)$ which is a composition of intrinsic motivation with integrated regulation items.

As can be observed in figure 1, the results showed a trend of alignment of the items with the self-determination continuum. Note that some items did not follow this tendency: 2, 8, 12, 14 and 23 are in different groups than expected by the questionnaire proposal. Item 2, that should be identified regulation was perceived as external regulation. This behavior, also observed in Guimarães and Bzuneck (2008) and Leal et al. (2013). It may have occurred due to the word "attendance" (see questionnaire in appendix), which led students to interpret the statement differently than expected. Item 8, proposed as introjected regulation, was perceived as external regulation in IB. For our sample item 12 (integrated regulation) and 14 (external regulation) were perceived as introjected regulation. Note that the analysis with IB item 23 (introjected regulation) is always in the same factor as 22 (identified regulation), even when the groups are dismembered. In Leal et al. (2013) analysis items 22 and 23 are observed in same factor. It can also be noticed a close proximity between amotivation with external regulation and between integrated regulation with intrinsic motivation. Analyzing gradual partition of items with IB, identified, integrated regulations and intrinsic motivation are in the same group as partition progresses. Identified, integrated, and intrinsic combining to form an autonomous motivation composite was observed by Ryan and Deci (2000).

We compared items partition produced by IB with the traditional methods hierarchical clustering and k-means with Euclidean distances, both using Statistica software (V8.0). The dichotomous distinction of **intrinsic and extrinsic motivation** items obtained with IB method is observed in the hierarchical dendrogram, shown in Fig. (2), with a difference for items 22 (identified regulation) and 23 (introjected regulation) that joins **extrinsic motivation** group, however with a weak connection indicated by a high linkage distance.

The k-means algorithm produced the same dichotomy, the only difference from IB was item 23 joining **extrinsic motivation group**. The persistent subset {2,3,8,11} obtained with IB that shows a striking difference from questionnaire proposal is also identified with hierarchical dendrogram and k-means. The last one joined item 29 (external regulation) to this set. Another result to mention is IB, hierarchical and k-means joining the intrinsic and integrated regulation.

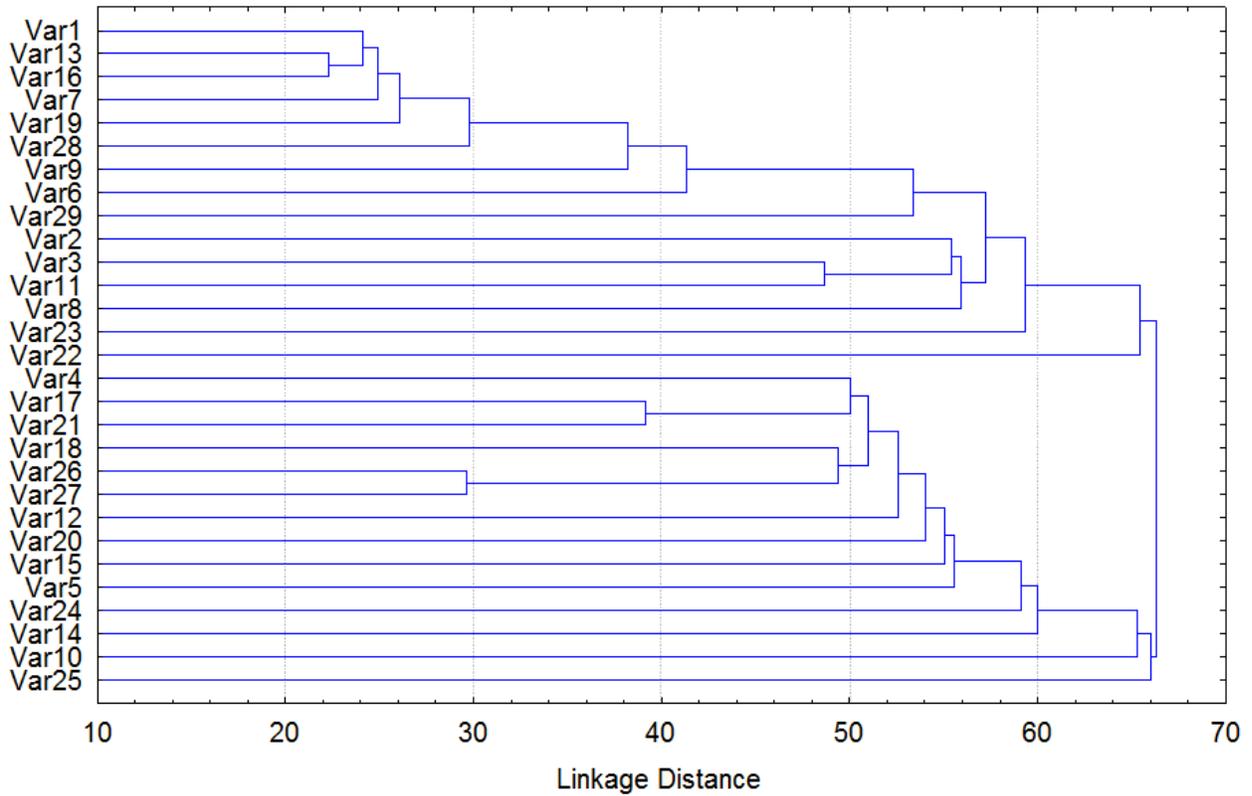

Figure 2: Dendrogram of hierarchical clustering of the questionnaire items.

IB method application does not involve assumptions about the sample, and has a process of identification of the most appropriate configuration by selecting the lowest functional shown in Eq. (2). All methods tested agree that questionnaire items 2, 14 and 23, were not perceived by the sample as expected, since they were not grouped according to this adapted instrument (appendix). These founds strongly indicate that some items of this instrument should be reformulated.

14## Conclusion

The results obtained in this work suggest that IB can be applied to the study of psychometric properties of instruments developed for the study of motivational profiles. Without the need to establish a metric, IB method compacts questionnaires items into smaller group configurations with the least loss of information about respondents guaranteed by minimizing a functional. Applied to an adapted instrument, it allowed through a gradual partition of the items, the identification of how the continuum of motivational states proposed by Deci and Ryan (2000) emerges, i.e. this method allowed to corroborate the gradual partition of items according to self-determination continuum.

The IB method was applied to an instrument developed for the study of the student motivation to attend the university, proposed by Guimarães and Bzuneck (2008), an adaptation of Vallerand AMS. The main contribution of these authors was to introduce new items to measure the integrated regulation. To Vansteenkiste, Lens and Deci (2006) the identified regulation was considered the most self-determined extrinsic motivation form, but for Fairchild et al. (2005) there is a need to contemplate the integrated regulation in new studies. Guimarães and Bzuneck (2008) obtained 7 factors, but some items were not perceived correctly, that is, were not in the expected factor. Results obtained by Leal et al. (2013) showed the same behavior and, in addition, integrated regulation and intrinsic motivation were in the same factor. Such results were corroborated with factor analysis and also with k-means and hierarchical clustering of our data sample.

The application of the IB to the questionnaire identified the problematic items directly without the need for arbitrary decisions made by the researcher such as identification of the scree plot elbow, attempts of different rotations, or decision on the number of factors based on the accumulated total variance.

Each possible items configuration is associated to a value of IB functional from Eq.2. Submitting to the functional the configuration that should have been found according to the questionnaire proposal (Guimarães & Bzuneck, 2008), the obtained value is higher than the obtained through IB iterative process, showing a higher loss of information about the respondents.

The main differences of items allocation into groups were also observed using EFA, k-means and hierarchical. Despite of IB clustering heuristics be an iterative approach that converges to a local minimum, the measure of efficiency based on the tradeoff between accuracy and complexity (Tishby, Pereira, and Bialek, 2000) allows inspection of solutions obtained by means of random initializations of the conditional probability of questionnaire items into clusters.

As a conclusion, the results of this first application of IB method to psychometric studies suggested that in conducting exploratory analyses the properties of the questionnaire be investigated to verify through items set partition, if the groups follow the trend of the self-determination continuum. IB could be useful as a consistent, complementary method to traditional exploratory procedures.

**Appendix**: Questionnaire items (Leal et al. 2013).

Amotivation

1- Honestly, I do not know why I come to the university

7- I really feel that I am wasting my time at the university

9- I had good reasons for coming to the university, but I now have doubts about continuing

13- I do not see why I must come to the university

16- I do not know or understand what I'm doing at the university

19- I do not see what difference it makes coming to the university

Introjected Regulation

5- I come to the university to prove to myself that I am able to complete the program

8 - I come because that is what is expected of me

10 - To prove to myself that I am an intelligent person

15 - I come to the university because being successful makes me feel important

20 - Because I want to prove to myself that I can be successful in my studies

23 - I want to avoid people seeing me as someone who has flunked out

External Regulation by Attendance

3 - I come to the university so that I will not fail

11 - I come to the university because attendance is mandatory

14 - I come to the university to obtain a diploma

External Regulation by Social Interaction

6 - I come to the university so I don't have to stay at home

28 - I come to the university because as long as I am studying, I do not have to work

29 - Seeing my friends is the main reason why I come to the university

Identified Regulation

2 - I come to the university because I think attendance should be mandatory

22 - Because I think attendance is required for students to take the program seriously

24 - I come to the university because attendance is necessary for learning

25 - If attendance was not mandatory, few students would attend classes

Integrated Regulation

12 - Because education is a privilege

18 - Because access to knowledge takes place at the university

26 - Because studying broadens one's horizons

27 - I come to the university because that is what I chose for myself

Intrinsic Motivation

4 - For the pleasure I obtain by engaging in interesting discussions with professors

17 - Because the university is a pleasure for me

21 - Because I love coming to the university